\documentclass[11pt]{article}
\usepackage{Blois,epsfig}

\def\gsim{\mathrel{\rlap{\lower4pt\hbox{$\sim$}}\raise1pt\hbox{$>$}}}

\def\be{\begin{equation}}
\def\ee{\end{equation}}
\def\bea{\begin{eqnarray}}
\def\eea{\end{eqnarray}}

\begin{document}
\vspace*{2cm}
\begin{center}
{\large \textit{pp}$\;$}{\large \textbf{Elastic Scattering in Near Forward Direction at LHC and Nucleon Structure}}
\end{center}
\begin{center}
{M.M. ISLAM$^*$ and R.J. LUDDY$^{\dagger}$\\
\small \textit{{Department of Physics, University of Connecticut, Storrs, CT 06269, USA}}}\\
\small \textit{{$^*$islam@phys.uconn.edu \hspace{1cm}$^{\dagger}$rjluddy@phys.uconn.edu}}\\
\vspace{.1cm}
{A.V. PROKUDIN\\
\small \textit{{Dipartmento di Fisica Teorica, Universit$\hat{a}$
Degli Studi di Torino, Via Pietro Giuria 1,\\
10125 Torino, Italy and Sezione INFN di Torino, Italy\\
Institute for High Energy Physics, 142281 Protvino, Russia\\
prokudin@to.infn.it}}}\\
(Presented by M. M. Islam)
\end{center}
\begin{abstract}
We predict \textit{pp} elastic differential cross section at LHC at the c.m. energy 
$\sqrt s $ = 14 TeV and momentum transfer range $\vert t\vert $ = 0 -- 10 
GeV$^{2}$, which is planned to be measured by the TOTEM group. The field 
theory model underlying our phenomenological investigation describes the 
nucleon as a composite object with an outer cloud of quark-antiquark 
condensate, an inner core of topological baryonic charge, and a still 
smaller quark-bag of valence quarks. The model satisfactorily describes the 
asymptotic behavior of $\sigma _{tot}$(s) and $\rho $(s) as well as the 
measured $\bar {p}p$ elastic d$\sigma $/dt at $\sqrt s $ = 546 GeV, 630 GeV, 
and 1.8 TeV. The large $\vert t\vert $ elastic amplitude of the model 
incorporates the QCD hard pomeron (BFKL Pomeron plus next to leading order 
approximations), the perturbative dimensional counting behavior, and the confinement 
of valence quarks in a small region within the nucleon.  Our predicted \textit{pp} 
elastic d$\sigma $/dt at LHC is compared with those of Bourrely et al. and 
Desgrolard et al.
\end{abstract}
As we all know, this is the 20th anniversary of the Blois Workshop.  So, delving a bit 
in history is only natural, and we find in the Proceedings of the First 
Blois Workshop, the editors -- Basarab Nicolescu and Tran Thanh Van -- quoted a 
foresightful observation by Elliot Leader in CERN Courier: ``..although asymptopia may 
be very far away indeed, the path to it is not through a desert but through a 
flourishing region of exciting physics...''  With the advent of the Large Hadron Collider 
(LHC) and the planned Total and Elastic Measurement (TOTEM) experiment, we are indeed 
entering a region of exciting physics -- the 10 TeV c.m. energy region.  TOTEM aims at 
measuring $\sigma _{tot}$, $\rho $ and d$\sigma $/dt up to $\vert t\vert $ $\simeq $ 10 GeV$^{2}$ 
at LHC at an unprecedented c.m. energy 14 TeV- an ambitious and challenging task.\cite{to}

Three groups have predicted \textit{pp} elastic differential cross section at LHC 
from $\vert t\vert $ = 0 all the way up to $\vert t\vert $ = 10 GeV$^{2}$ on the basis 
of three different models: 1) impact-picture model based on the Cheng-Wu calculations 
of QED tower diagrams;\cite{bsw} 2) eikonalized pomeron-reggeon model using conventional 
regge pole approach, but with multiple pomeron, reggeon exchanges included ;\cite{des} 
3)nucleon-structure model where the nucleon has an outer cloud of quark-antiquark 
condensed ground state, an inner core of topological baryonic charge, and a still 
smaller quark-bag of valence quarks (Fig.1).\cite{mm1,mm2}  A QCD-inspired eikonalized model 
has also been proposed\cite{bl} to predict \textit{pp} d$\sigma $/dt at $\sqrt s $ = 14 TeV 
for $\vert t \vert $ = 0 -- 2.0 GeV$^{2}$.

Our initial investigation led us to the following description.  The nucleon has an 
outer cloud and an inner core. High energy elastic scattering is primarily due to 
two processes: a) a glancing collision where the outer cloud of one nucleon interacts 
with that of the other giving rise to diffraction scattering; b) a hard, or large 
$\vert t \vert $ collision where one nucleon core scatters off the other core via 
vector meson $\omega $ exchange, while their outer clouds overlap and interact 
independently.  In the small $\vert t \vert $ region diffraction dominates, but the hard 
scattering takes over as $\vert t \vert $ increases.

We describe diffraction scattering using the impact parameter representation:

\begin{flushright}
$T_D (s,t)=i\,p\,W\int_0^\infty {b\;db\;J_0 } (b\,q)\Gamma_{D}(s,b)$;\hspace{4cm} (1)\\
\end{flushright}
here $\Gamma_{D}(s,b)$ is the profile function, which is related to the eikonal 
function $\chi_{D}(s,b)$:
$\Gamma_{D}(s,b) = 1 - exp(i \chi_{D}(s,b))$. We choose $\Gamma_{D}(s,b)$ to
be an even Fermi profile function:

\begin{flushright}
$\Gamma_{D}(s,b) = g(s)$ [$ 1 \over 1 + exp((b-R)/a)$ + $ 1 \over 1 + exp(-(b+R)/a)$ -1 ].\hspace{3 cm}(2)\\
\end{flushright}

Besides $g(s)$, the parameters $R$ and $a$ are also energy dependent. Further 
studies show that $R$ and $a$ have the following energy dependences : 
$R$ = $R_0 + R_1(ln s -$ $i\pi\over 2$ $)$,
$a$ = $a_0 + a_1(ln s -$ $i\pi\over 2$ $)$;
 $g(s)$ is a complex crossing even energy-dependent function that asymptotically becomes 
a real positive constant.  The diffraction amplitude we obtain satisfies a number 
of general properties associated with the phenomenon of diffraction:
\vspace{.2cm}\\
1.  $\sigma_{tot}(s) \sim (a_{0} + a_{1} ln s)^2$ (Froissart-Martin bound)\\
2.  $\rho (s) \simeq \frac{\pi a_1}{a_0 + a_1 ln s}$ (derivative dispersion relation)\\
3.  $T_{D}(s,t) \sim i\; s\; ln^{2}s\; f(|t| ln^{2}s)$  (AKM scaling)\\
4.  $T_{D}^{\bar{p}p}(s,t) = T_{D}^{pp}(s,t)$      (crossing even).
\vspace{.2cm}\\
Incidentally, the profile function (2) has been used by Frankfurt et al.\cite{fr} to represent 
the impact parameter distribution of soft inelastic collisions in diffractive Higgs production at LHC.

We take the hard scattering amplitude due to $\omega $ exchange to be of the form
\begin{flushright}
$T^{H}_{\omega}(s,t)\;\sim \;\exp [i\;\hat{\chi }(s,0)]\;s\;\frac{F^2(t)}{m_\omega ^2 -t}$.\hspace{4.3cm}(3)\\
\end{flushright}
The t-dependence is the product of two form factors and
the $\omega$ propagator.  It shows that $\omega $ probes two density
distributions corresponding to the two form factors. The density
distributions represent the nucleon cores.  The factor of s originates
from spin 1 of $\omega$.  Such an s-dependence is not expected in a regge pole 
model, but can occur in the nonlinear $\sigma $ model where $\omega $ couples to 
the baryonic current and the baryonic current is topological.

Our phenomenological investigation progressively led us to an effective field 
theory model--a gauged linear $\sigma $ model of the Gell-Mann-Levy type, 
and the physical structure of the nucleon that emerges is that shown 
in Fig.1.\cite{mm2}  This structure indicates that at very large 
$\vert t \vert $ and therefore small $b$, the quark-bag of one nucleon overlaps 
that of the other and large $\vert t \vert $ elastic scattering originates from 
valence quark-quark scattering.  We view this process as shown in Fig. 2.  
A valence quark from one proton makes a hard, i.e. a large $\vert t \vert $ 
collision with a valence quark from the other proton.  The collision carries 
off the whole momentum transfer.  This dynamical picture brings new features 
into our calculations: 1) probability amplitude of a quark to have, say, 
momentum $\vec{p}$ when the proton has momentum $\vec{P}$ in the c.m. frame; 
2) quark-quark elastic amplitude at high energy and large momentum transfer, 
which we take as the BFKL pomeron with next to leading order corrections included.\cite{mm1} 
The latter amplitude is referred by us as the QCD hard pomeron.

To obtain the \textit{pp} elastic amplitude $T_{qq}(s,t)$ due to quark-quark scattering 
as depicted in Fig. 2, we have to introduce two ``structure factors'' that take into 
account the momentum wave function of each valence quark in a proton and the 
fractional longitudinal momentum it carries.  The net result is that we now 
have a second hard amplitude of the form
\begin{flushright}
T$^{H}_{qq}(s,t) \sim i\;s\;\exp [i\;\hat{\chi }(s,0)]\,(s\;e^{-i\frac{\pi}{2}})^\omega 
\;\frac{\mathcal{F}^{2}(q_\bot )}{\vert t\vert +r_0^{-2}}$ , \hspace{3.4cm}(4)\\
\end{flushright}
where $\alpha_{BFKL}=1 + \omega$ and $r_{0}$ defines the black disk radius of
asymptotic quark-quark scattering.  $\mathcal{F}(q_{\bot})$ is the structure factor 
which should be distinguished from a form factor.  Our study of large $q^{2}_{\bot}$ 
behavior of $\mathcal{F}(q_{\bot})$ leads to $T_{qq}(s,t)/s \sim \vert t \vert ^{-5}$ 
and results in a differential cross section behavior 
d$\sigma $/dt $\sim \vert t \vert ^{-10}$ as predicted by perturbative QCD quark 
counting rules for fixed $s$ and large $\vert t \vert $ ($s>>\vert t >>M^{2}$).

We determine the parameters of the model by requiring that the model should 
describe satisfactorily the asymptotic behavior of $\sigma _{tot}$(s) and  
$\rho $(s) as well as the experimentally measured $\bar{p}p$ elastic 
d$\sigma $/dt at $\sqrt{s}$ = 546 GeV, 630 GeV and 1.8 TeV.  The results of 
this investigation for $\sigma _{tot}$(s) and $\rho $(s) are given in Figs. 3 and 4, 
where the dotted curves represent the error bands given by Cudell et al. (COMPETE 
Collaboration) to their best fit.\cite{cu} Our prediction for $pp$ elastic differential 
cross section at LHC at $\sqrt{s}$ = 14 TeV for $\vert t \vert $ = 0 -- 10 GeV$^{2}$ 
is given in Fig. 5.  Also shown in this figure are separate d$\sigma $/dt due to 
diffraction (dotted curve), due to hard $\omega $ exchange (dot-dashed curve), and 
due to hard $qq$ scattering (dashed curve).  As we mentioned earlier, two other groups, 
Bourrely et al.\cite{bsw} and Desgrolard et al.\cite{des} have predicted $pp$ elastic d$\sigma $/dt 
at LHC from $\vert t \vert $ = 0 -- 10 GeV$^{2}$.  In Fig. 6, a comparison of our 
prediction (solid curve) with those of Bourrely et al. (dot-dashed curve) and 
Desgrolard et al. (dashed curve) are given.  We notice that while we predict smooth 
fall-off of d$\sigma $/dt for $\vert t \vert \gsim $ 1 GeV$^{2}$, the latter authors predict 
oscillations.  Furthermore, we predict much larger differential cross section 
at large $\vert t \vert $ than them.

We conclude by noting that precise measurement of $pp$ elastic d$\sigma $/dt 
by the TOTEM group at LHC at c.m. energy 14 TeV and $\vert t \vert $ = 0 -- 10 GeV$^{2}$ 
will be able to verify the composite structure of the nucleon and the QCD hard pomeron 
contribution, which have emerged from our investigation.

\begin{figure}[b]
  \vspace{-7cm}
  \center{
  \includegraphics[height=2.7in]{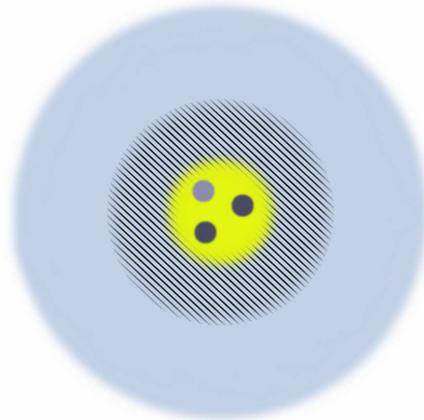}
  \caption{Nucleon structure emerging from our investigation.  Nucleon has an outer cloud of $q\bar{q}$ condensed ground state analogous to the BCS ground state in superconductivity, an inner core of topological baryonic charge probed by $\omega $, and a still smaller quark-bag of massless valence quarks.}
  }
\end{figure}
\begin{figure}[t]
  \center{
  \includegraphics[height=1.5in]{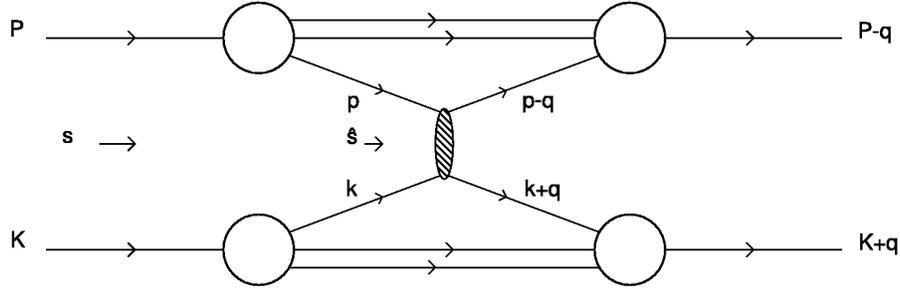}
  \vspace{-0.4cm}
  \caption{Hard collision of valence quarks from two different protons.}
  }
\end{figure}
\begin{figure}[b]
\centering
\includegraphics[height=2.7in]{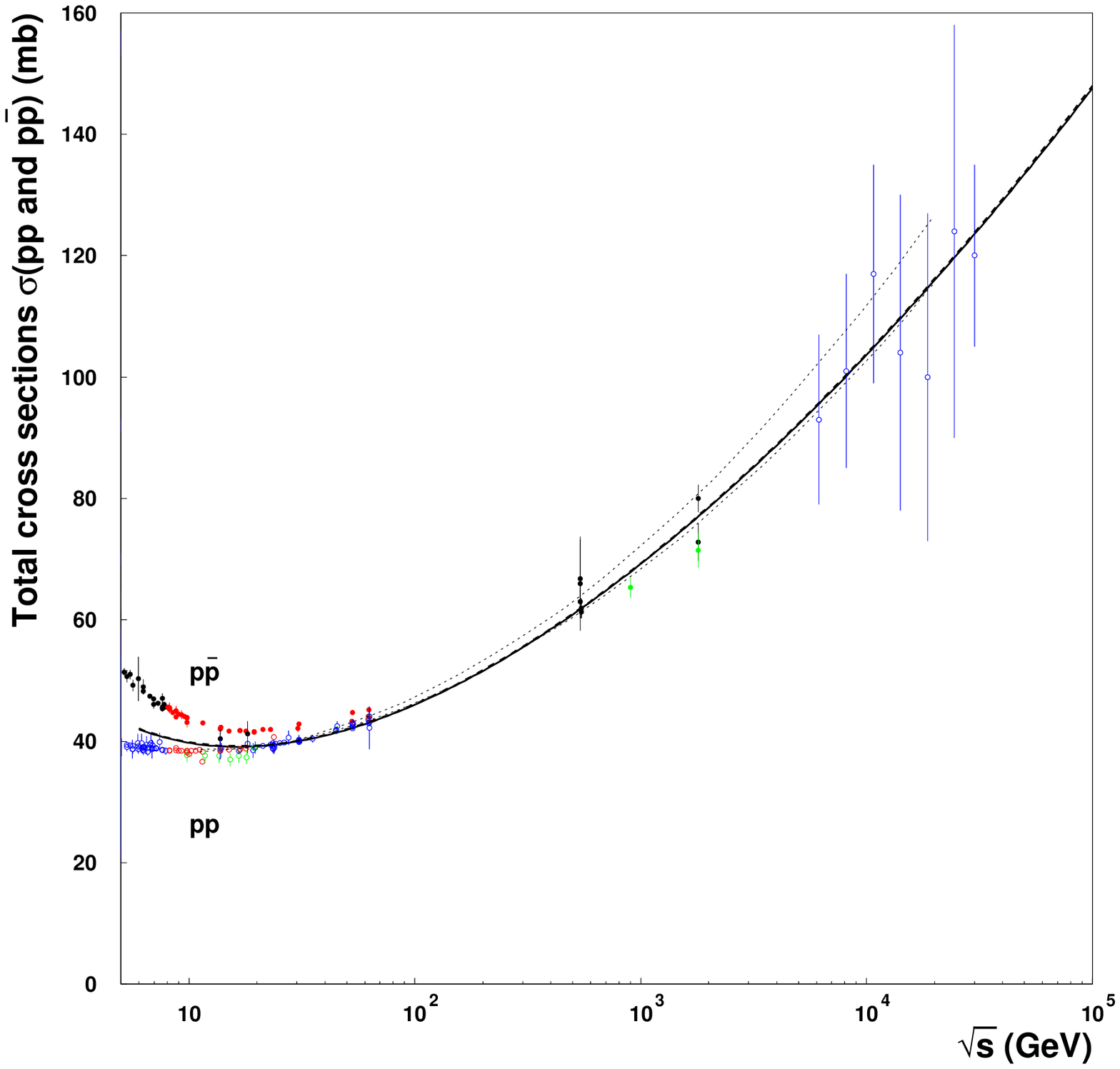}
\hspace{1cm}
\includegraphics[height=2.7in]{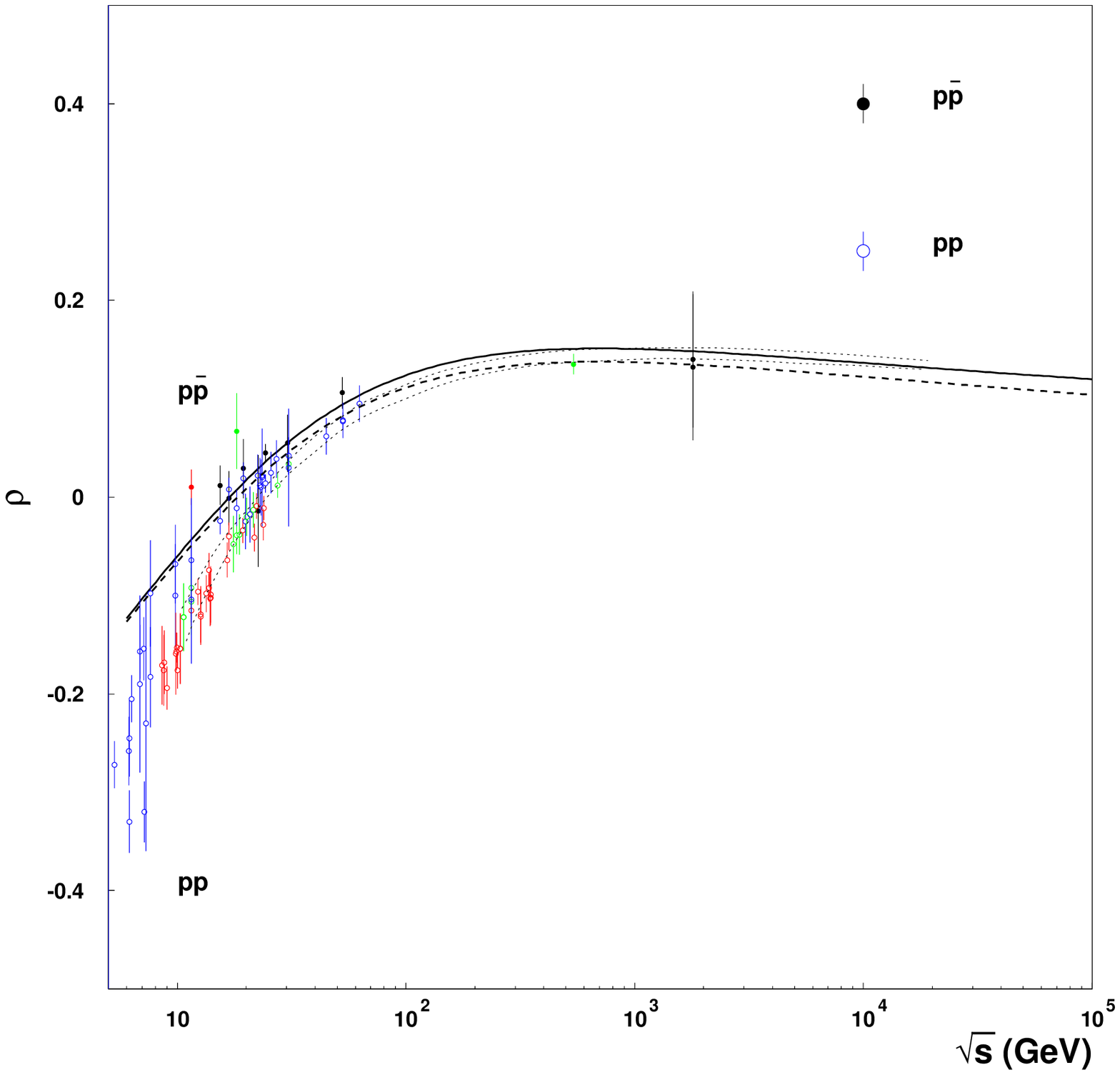}
\vspace{-0.4cm}
\caption{
Solid curve represents our calculated\hspace{2.04cm}Figure 4: Solid and dashed curves represent our\hspace{0.3cm}$\;$\newline
\hspace{0.7cm}total cross section as a function of $\sqrt{s}$. Dotted \hspace{2.cm}calculated $\rho_{\bar{p}p}$ and $\rho_{pp}$ as functions of $\sqrt{s}$. Dot-\hspace{0.3cm}$\;$\newline
curves $\;$represent$\;$ the$\;$ error$\;$ band$\;$ given by \hspace{2.4cm}ted curves $\;$represent $\;$the $\;$error $\;$band $\;$given by\hspace{0.3cm}$\;$\newline 
Cudell et al.\protect\cite{cu}\hspace{6.88cm}Cudell et al.\protect\cite{cu}\hspace{5.12cm}$\;$
}
\end{figure}
\setcounter{figure}{4}
\begin{figure}[b]
\centering
\includegraphics[height=2.7in]{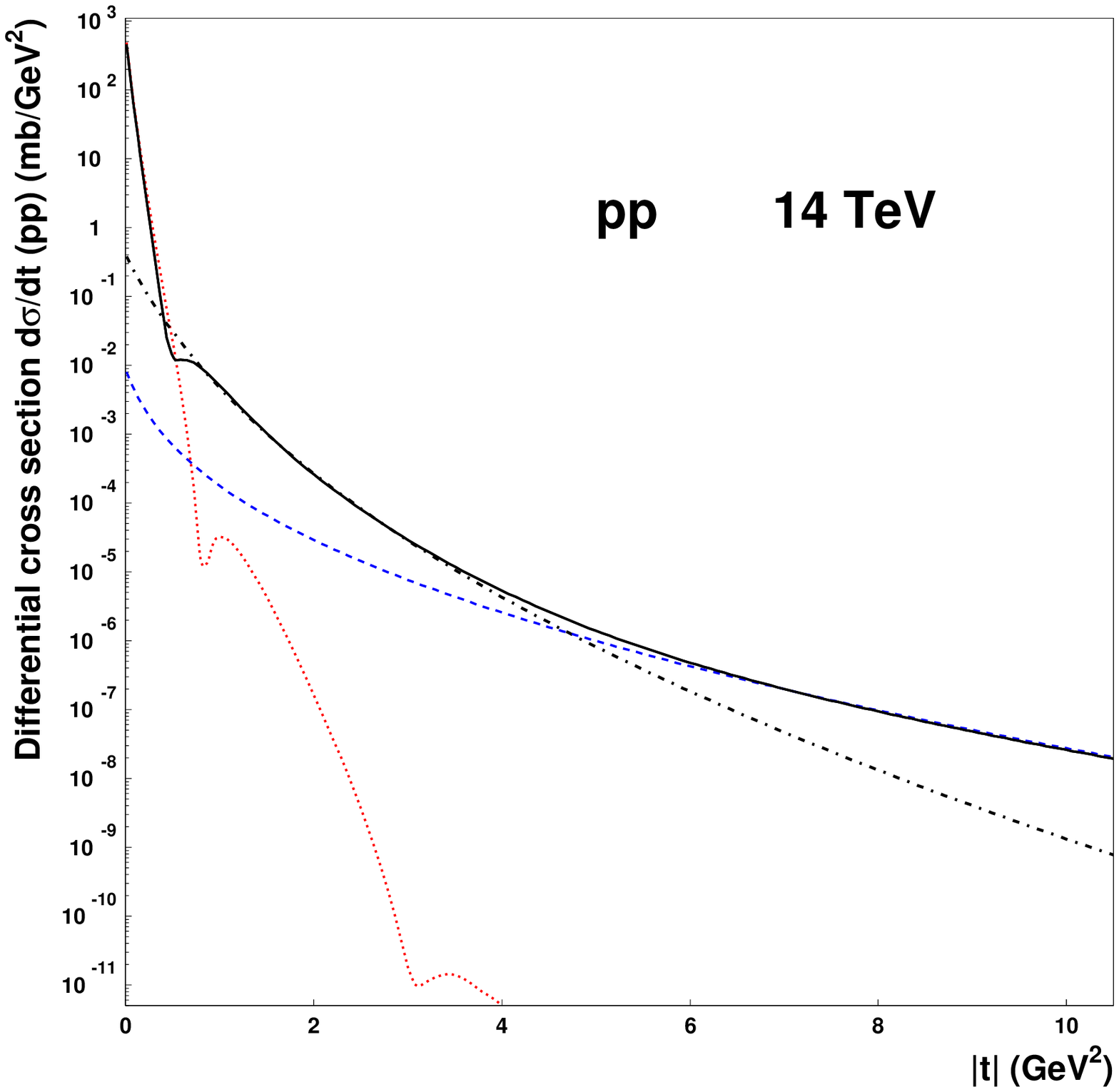}
\hspace{1cm}
\includegraphics[height=2.7in]{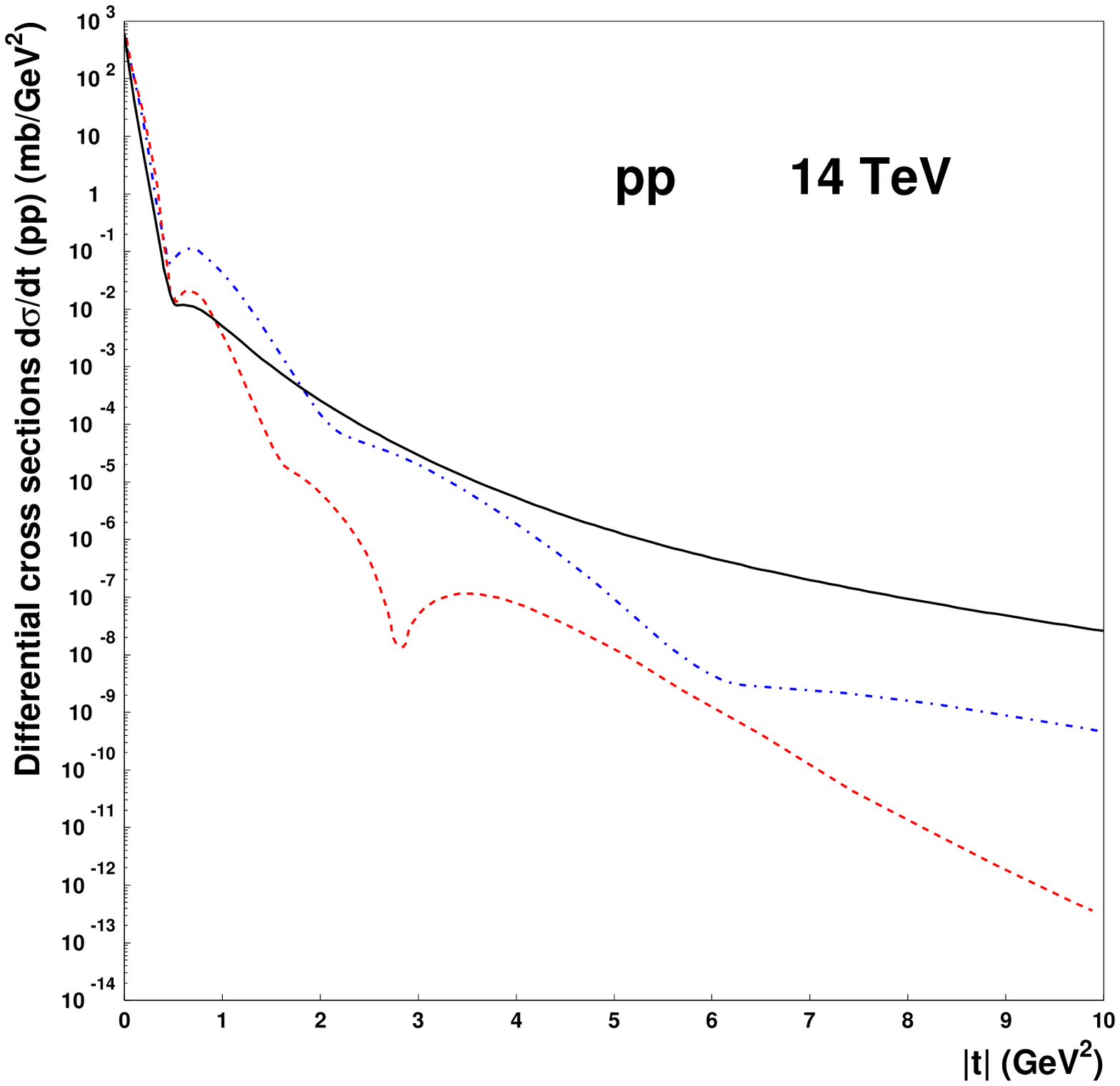} 
\vspace{-0.4cm}
\caption{
Solid curve shows our predicted d$\sigma $/dt \hspace{1.5cm}Figure 6: Comparison of our predicted differential\newline
for $pp$ elastic scattering at $\sqrt{s}$=14 TeV at LHC.\hspace{1.7cm}cross section (solid curve)$\;$ with those of Bourrely$\;$\newline
Dotted curve represents d$\sigma $/dt due to diffraction \hspace{1.5cm}et al.$\;$ (dot-dashed$\;$ curve) $\;$and$\;$ Desgrolard$\;$ et al.$\;$\newline
only.  Similarly, dot-dashed and dashed curves \hspace{1.8cm}(dashed curve).\hspace{4.9cm}$\;$\newline
represent d$\sigma $/dt due to hard $\omega $-exchange and\hspace{9.2cm}$\;$\newline
hard $qq$ scattering only.\hspace{11.83cm}$\;$
}
\end{figure}
\end{document}